\documentclass[a4,11pt]{article}
\usepackage{coling10}
\usepackage{color}
\usepackage{times,graphicx,latexsym}
\usepackage{booktabs}
\usepackage{mymath}

\graphicspath{{./}}
\title{Local Space-Time Smoothing for Version Controlled Documents}
\author{Seungyeon Kim\\
  Georgia Institute of Technology\\
\And
  Guy Lebanon\\
  Georgia Institute of Technology\\
 }
\begin{document}
\maketitle
\begin{abstract}
Unlike static documents, version controlled documents are continuously edited by one or more authors. Such collaborative revision process makes traditional modeling and visualization techniques inappropriate. In this paper we propose a new representation based on local space-time smoothing that captures important revision patterns. We demonstrate the applicability of our framework using experiments on synthetic and real-world data.
\end{abstract}

\section{Introduction} \label{sec:introduction}
Most computational linguistics studies concentrate on modeling or analyzing documents as sequences of words. In this paper we consider modeling and visualizing version controlled documents which is the authoring process leading to the final word sequence. In particular, we focus on documents whose authoring process naturally segments into consecutive versions. The revisions, as the differences between consecutive versions are often called, may be authored by a single author or by multiple authors working collaboratively.

One popular way to keep track of version controlled documents is using a version control system such as CVS or Subversion (SVN). This is often the case with books or with large computer code projects. In other cases, more specialized computational infrastructure may be available, as is the case with the authoring API of Wikipedia.org, Slashdot.com, and Google Wave. Accessing such API provides information about what each revision contains, when was it submitted, and who edited it. In any case, we formally consider a version controlled document as a sequence of documents $d_1,\ldots,d_l$ indexed by their revision number where $d_i$ typically contains some locally concentrated additions or deletions, as compared to $d_{i-1}$.

In this paper we develop a continuous representation of version controlled documents that generalizes the locally weighted bag of words representation \cite{Lebanon2007b}. The representation smooths the sequence of version controlled documents across two axes-time $t$ and space $s$. The time axis $t$ represents the revision and the space axis $s$ represents document position. The smoothing results in a continuous map from a space-time domain to the simplex of term frequency vectors
\begin{align} \label{eq:simplex}
\gamma&:\Omega \to \mathbb{P}_V \quad \text{where} \quad \Omega\subset \mathbb{R}^2, \quad \text{and}\\
\mathbb{P}_V &=\left\{ w\in\mathbb{R}^{|V|}: w_i\geq 0, \,\,\, \sum_{i=1}^{|V|} w_i=1\right\}.  \nonumber
\end{align}
The mapping above ($V$ is the vocabulary) captures the variation in the local distribution of word content across time and space. Thus $[\gamma(s,t)]_w$ is the (smoothed) probability of observing word $w$ in space $s$ (document position) and time $t$ (version). Geometrically, $\gamma$ realizes a divergence-free vector field (since $\sum_w [\gamma(s,t)]_w=1$, $\gamma$ has zero divergence) over the space-time domain $\Omega$.

We consider the following four version controlled document analysis tasks. The first task is visualizing word-content changes with respect to space (how quickly the document changes its content), time (how much does the current version differs from the previous one), or mixed space-time. The second task is detecting sharp transitions or edges in  word content. The third task is concerned with segmenting the space-time domain into a finite partition reflecting word content. The fourth task is predicting future revisions. Our main tool in addressing tasks 1-4 above is to analyze the values of the vector field $\gamma$ and its first order derivatives fields \begin{align}\label{eq:gradient}
\nabla \gamma&=\left(\dot{\gamma}_s,\dot{\gamma}_t\right).
\end{align}

\section{Space-Time Smoothing for Version Controlled Documents}
\label{spacetime}
With no loss of generality we identify the vocabulary $V$ with positive integers $\{1,\ldots,V\}$ and represent a word $w\in V$ by a unit vector\footnote{Note the slight abuse of notation as $V$ represents both a set of words and an integer $V=\{1,\ldots,V\}$ with $V=|V|$.} (all zero except for 1 at the $w$-component)
\begin{align} \label{eq:unitVector}
e(w)=(0,\ldots,0,1,0,\ldots,0)^{\top}\quad w\in V.
\end{align}
We extend this definition to word sequences thus representing documents $\langle w_1,\ldots,w_N\rangle$ ($w_i\in V$) as sequences of $V$-dimensional vectors $\langle e(w_1),\ldots,e(w_N)\rangle$. Similarly, a version controlled document is sequence of documents $d^{(1)},\ldots,d^{(l)}$ of potentially different lengths $d^{(j)}=\langle w_1^{(j)},\ldots,w_{N(j)}^{(j)}\rangle$. Using \eqref{eq:unitVector} we represent a version controlled document as the array
\begin{align} \label{eq:raggedArray}
\begin{array}{lcr}
e(w_1^{(1)}), &\ldots, &e(w_{N(1)}^{(1)})\\
\vdots        &\ddots &\vdots\\
e(w_1^{(l)}), &\ldots, &e(w_{N(l)}^{(l)})
\end{array}
\end{align}
where columns and rows correspond to space (document position) and time (versions).

The array \eqref{eq:raggedArray} of high dimensional vectors represents the version controlled document without any loss of information. Nevertheless the high dimensionality of $V$ suggests we smooth the vectors in \eqref{eq:raggedArray} with neighboring vectors in order to better capture the local word content. Specifically we convolve each component of \eqref{eq:raggedArray} with a 2-D smoothing kernel  $K_h$ to obtain a smooth vector field $\gamma$ over space-time \cite{Wand1995} e.g.,
\begin{align}\nonumber
 \label{eq:gamma}
\gamma(s,t) &= \sum_{s'}\sum_{t'} K_h(s-s',t-t') e(w_{s'}^{(t')})\\
K_h(x,y)&\propto \exp\left(-(x^2+y^2)/(2h^2)\right).
\end{align}
Thus as $(s,t)$ vary over a continuous domain $\Omega\subset\mathbb{R}^2$, $\gamma(s,t)$, which is a weighted combination of neighboring unit vectors, traces a continuous surface in $\mathbb{P}_V\subset \mathbb{R}^V$. Assuming that the kernel $K_h$ is a normalized density it can be shown that $\gamma(s,t)$ is a non-negative normalized vector i.e., $\gamma(s,t)\in\mathbb{P}_V$ (see \eqref{eq:simplex} for a definition of $\mathbb{P}_V$) measuring the local distribution of words around the space-time location $(s,t)$. It thus extends the concept of lowbow (locally weighted bag of words) introduced in \cite{Lebanon2007b} from single documents to version controlled documents.

One difficulty with the above scheme is that the document versions $d_1,\ldots,d_l$ may be of different lengths. We consider two ways to resolve this issue. The first pads  shorter document versions with zero vectors as needed. We refer to the resulting representation $\gamma$ as the non-normalized representation. The second approach normalizes all document versions to a common length, say $\prod_{j=1}^l N(j)$. That is each word in the first document is expanded into $\prod_{j\neq 1} N(j)$ words, each word in the second document is expanded into $\prod_{j\neq 2} N(j)$ words etc. We refer to the resulting representation $\gamma$ as the normalized representation.

\begin{figure*} \centering
\includegraphics[width=0.23\textwidth]{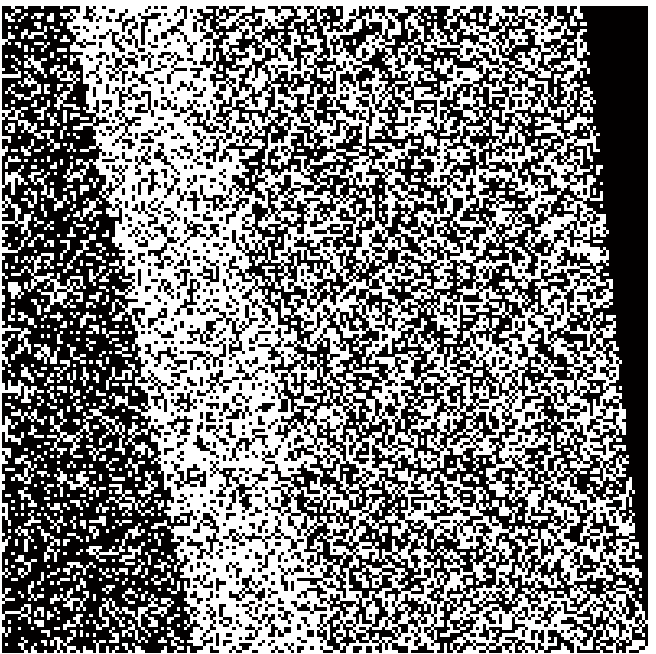}
\includegraphics[width=0.23\textwidth]{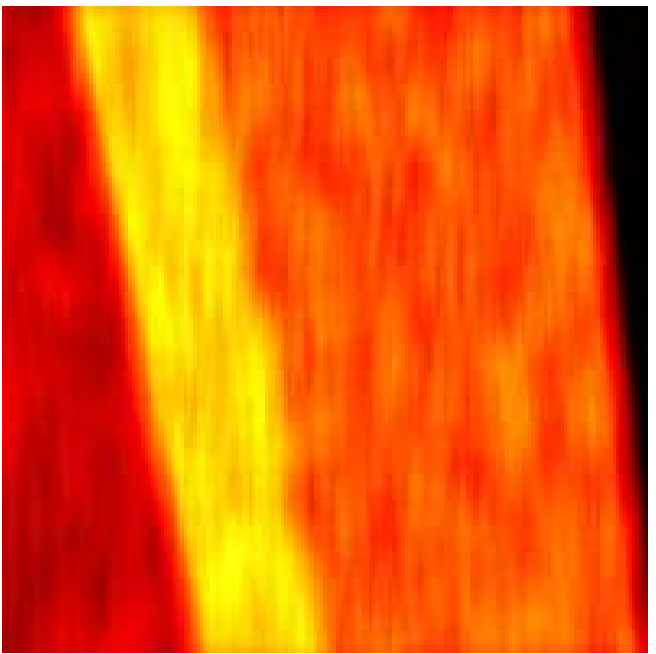}
\includegraphics[width=0.23\textwidth]{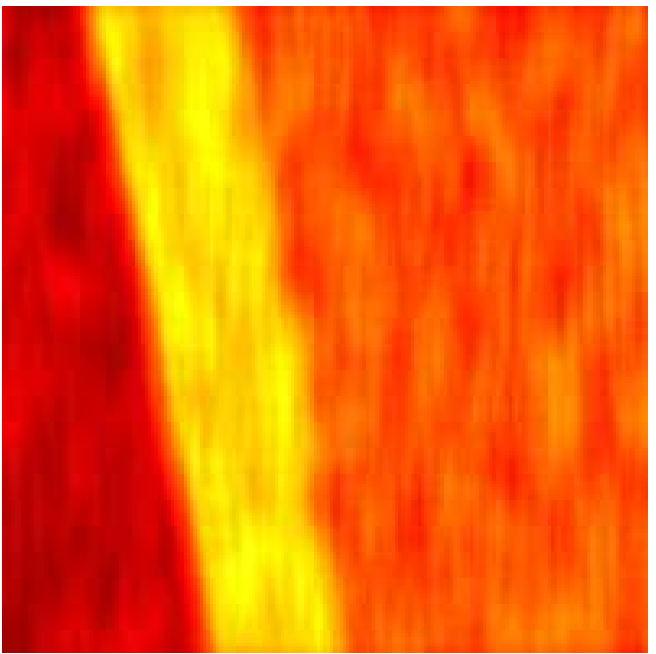}
\scalebox{1}[1.27]{\includegraphics[width=0.23\textwidth]{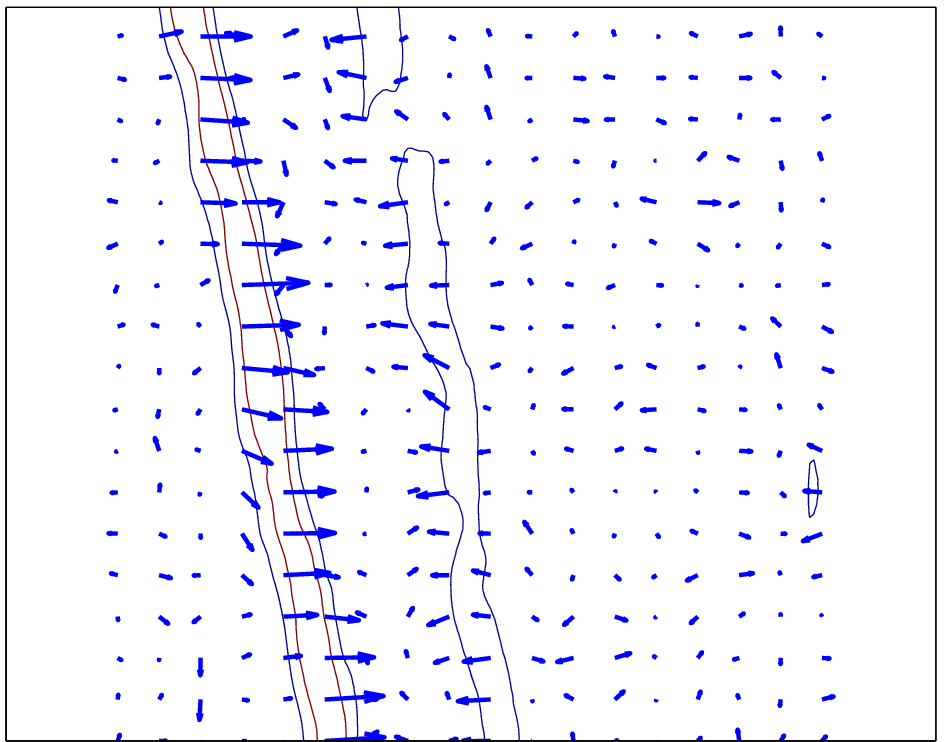}}
\caption{ \small Four space-time representations of a simple synthetic version controlled document over $V=\{1,2\}$ (see text for more details). The left panel displays the first component of \eqref{eq:raggedArray} (non-smoothed array of unit vectors corresponding to words). The second and third panels display $[\gamma(s,t)]_1$ for the non-normalized and normalized representations respectively. The fourth panel displays the gradient vector field $(\dot\gamma_s(s,t),\dot\gamma_t(s,t))$ (contour levels represent the gradient magnitude). The black portions of the first two panels correspond to zero padding due to unequal lengths of the different versions.}   \label{fig:synFigLebanon}
\end{figure*}

The non-normalized representation has the advantage of conveying absolute lengths. For example, it makes it possible to track how different portions of the document grow or shrink (in terms of number of words) with the version number. The normalized representation has the advantage of conveying lengths relative to the document length. For example, it makes it possible to track how different portions of the document grow or shrink with the version number relative to the total document length. In either case, the space-time domain $\Omega$ on which $\gamma$ is defined \eqref{eq:gamma} is a two dimensional rectangular domain $\Omega=[0,I]\times[0,J]$.

Before proceeding to examine how $\gamma$ may be used in the four tasks described in Section~\ref{sec:introduction} we demonstrate our framework with a simple low dimensional example. Assuming a vocabulary of two words $V=\{1,2\}$ we can visualize $\gamma$ by displaying its first component as a grayscale image (since $[\gamma(s,t)]_2=1-[\gamma(s,t)]_1$ the second component is redundant). Specifically, we created a version controlled document with three contiguous segments whose $\{1,2\}$ words were sampled from Bernoulli distributions with parameters 0.3 (first segment), 0.7 (second segment), and 0.5 (third segment). That is, the probability of getting 1 is highest for the second segment, equal for the third and lowest for the first segment. The initial lengths of the segments were  30, 40 and 120 words with the first segment increasing and the third segment decreasing at half the rate of the first segment with each revision. The length of the second segment was constant across the different versions. Figure~\ref{fig:synFigLebanon} displays the nonsmoothed ragged array \eqref{eq:raggedArray} (left), the non-normalized $[\gamma(s,t)]_1$ (middle left) and the normalized $[\gamma(s,t)]_1$ (middle right).

\begin{figure*} \centering
\setlength\fboxsep{0pt} \setlength\fboxrule{0.5pt}
\fbox{\includegraphics[width=0.23\textwidth]{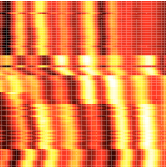}}
\fbox{\includegraphics[width=0.23\textwidth]{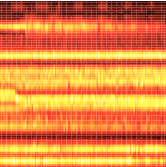}}
\fbox{\includegraphics[width=0.23\textwidth]{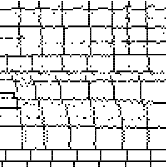}}
\fbox{\includegraphics[width=0.23\textwidth, height=0.23\textwidth]{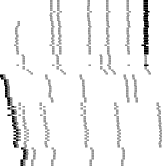}}
\caption{ \small Gradient and edges for a portion of the version controlled Wikipedia Religion article. The left panel displays $\|\dot\gamma_s(s,t)\|^2$  (amount of change across document locations for different versions). The second panel displays $\|\dot\gamma_t(s,t)\|^2$ (amount of change across versions for different document positions).  The third panel displays the local maxima of $\|\dot\gamma_s(s,t)\|^2+\|\dot\gamma_t(s,t)\|^2$ which correspond to potential edges, either vertical lines (section and subsection boundaries) or horizontal lines (between substantial revisions). The fourth panel displays boundaries of sections and subsections as black and gray lines respectively.} \label{fig:religionPartVis}
\end{figure*}

While the left panel doesn't distinguish much between the second and third segment the two smoothed representations display a nice segmentation of the space-time domain into three segments, each with roughly uniform values. The non-normalized representation (middle left) makes it easy to see that the total length of the version controlled document is increasing but it is not easy to judge what happens to the relative sizes of the three segments. The normalized representation (middle right) makes it easy to see that the first segment increases in size, the second is constant, and the third decreases in size. It is also possible to notice that the growth rate of the first segment is higher than the decay rate of the third.

\section{Visualizing Change in Space-Time}\label{sec:visChange}
We apply the space-time representation to four tasks. The first task, visualizing change, is described in this section. The remaining three tasks are described in the next three section.

The space-time domain $\Omega$ represents the union of all document versions and all document positions. Some parts of $\Omega$ are more homogeneous and some are less in terms of their local word distribution. Locations in $\Omega$ where the local word distribution substantially diverges from its neighbors correspond to sharp content transitions. On the other hand, locations whose word distribution is more or less constant correspond to slow content variation.

We distinguish between three different types of changes. The first occurs when the word content changes substantially between neighboring document positions within a certain document version. As an example consider a document location whose content shifts from high level introductory motivation to a detailed technical description. Such change is represented by
\begin{align} \label{eq:D1space}
\|\dot\gamma_s(s,t)\|^2 &= \sum_{w=1}^V \left(\frac{\partial[\gamma(s,t)]_w}{\partial s}\right)^2.
\end{align}

A second type of change occurs when a certain document position undergoes substantial change in local word distribution across   neighboring versions. An example is erroneous content in one version
being heavily revised in the next version. Such change along the time axis corresponds to the magnitude of
\begin{align} \label{eq:D1time}
\|\dot\gamma_t(s,t)\|^2 &= \sum_{w=1}^V \left(\frac{\partial[\gamma(s,t)]_w}{\partial t}\right)^2.
\end{align}

Expression \eqref{eq:D1space} may be used to measure the instantaneous rate of change in the local word distribution. Alternatively, integrating \eqref{eq:D1space} provides a global measure of change
\begin{align*}
h(s)=\int \|\dot\gamma_s(s,t)\|^2 \, dt, \quad
g(t)=\int \|\dot\gamma_t(s,t)\|^2 \, ds
\end{align*}
with $h(s)$ describing the total amount of spatial change across all revisions and $g(t)$ describing the total amount of version change across different document positions. $h(s)$ may be used to detect document regions undergoing repeated substantial content revisions and $g(t)$ may be used to detect revisions in which substantial content has been modified across the entire document.

We conclude with the integrated directional derivative
\begin{align} \label{eq:Dmixed}
\int_0^1 \| \dot\alpha_s(r)\dot\gamma_s(\alpha(r))  +   \dot\alpha_t(r)\dot\gamma_t(\alpha(r))       \|^2 \, dr
\end{align}
where $\alpha:[0,1]\to\Omega$ is a parameterized curve in the space-time and $\dot\alpha$ its tangent vector. Expression~\eqref{eq:Dmixed} may be used to measure change along a dynamically moving document anchor such as the boundary between two book chapters. The space coordinate of such anchor shifts with the version number (due to the addition and removal of content across versions) and so integrating the gradient across one of the two axis as in \eqref{eq:D1time} is not appropriate. Defining $\alpha(r)$ to be a parameterized curve in space-time realizing the anchor positions $(s,t)\in\Omega$ across multiple revisions, \eqref{eq:Dmixed} measures the amount of change at the anchor point.

\subsection{Experiments}
The right panel of Figure~\ref{fig:synFigLebanon} shows the gradient vector field corresponding to the synthetic version controlled document described in the previous section. As expected, it tends to be orthogonal to the segment boundaries. Its magnitude is displayed by the contour lines which show highest magnitudes around segment boundaries.

Figure~\ref{fig:religionPartVis} shows the norm  $\|\dot\gamma_s(s,t)\|^2$ (left), $\|\dot\gamma_t(s,t)\|^2$ (middle left) and the local maxima of $\|\dot\gamma_s(s,t)\|^2+\|\dot\gamma_t(s,t)\|^2$ (middle right) for a portion of the version controlled Wikipedia Religion article. The first panel shows the amount of change in local word distribution within documents. High values correspond to boundaries between sections, topics or other document segments. The second panel shows the amount of change as one version is replaced with another. It shows which revisions change the word distributions substantially and which result in a relatively minor change. The third panel shows only the local maxima which correspond to edges between topics or segments (vertical lines) or revisions (horizontal lines).

\begin{figure*} \centering
\setlength\fboxsep{0pt} \setlength\fboxrule{0.5pt}
\hspace{-3px}\fbox{\includegraphics[width=0.29\textwidth,height=0.29\textwidth]{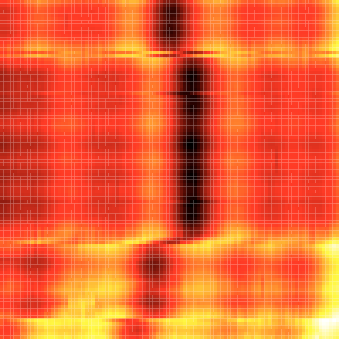}}
\fbox{\includegraphics[width=0.29\textwidth,height=0.29\textwidth]{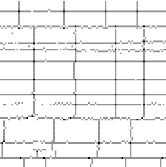}}
\fbox{\includegraphics[width=0.29\textwidth,height=0.29\textwidth]{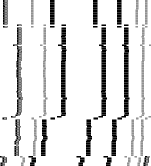}}
\fbox{\includegraphics[width=0.29\textwidth,height=0.29\textwidth]{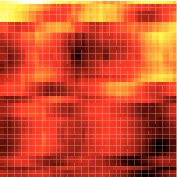}}
\fbox{\includegraphics[width=0.29\textwidth,height=0.29\textwidth]{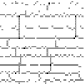}}
\fbox{\includegraphics[width=0.29\textwidth,height=0.29\textwidth]{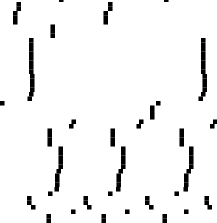}}
\vspace{-0.1in}
\caption{ \small Gradient and edges of a portion of the version controlled Atlanta Wikipedia article (top row) and the Google Wave Amazon Kindle FAQ (bottom row). The left column displays the magnitude of the gradient in both space and time $\|\dot\gamma_s(s,t)\|^2+\|\dot\gamma_t(s,t)\|$. The middle column displays the local maxima of the gradient magnitude (left column). The right column displays the actual segment boundaries as vertical lines (section headings for Wikipedia and author change in Google Wave). The gradient maxima corresponding to vertical lines in the middle column matches nicely the Wikipedia section boundaries. The gradient maxima corresponding to horizontal lines in the middle column correspond nicely to major revisions indicated by a discontinuities in the location of the section boundaries.} \label{fig:edgeVis}
\vspace{-0.13in}
\end{figure*}

\section{Edge Detection}
In many cases documents may be divided to semantically coherent segments. Examples of text segments include individual news stories in streaming broadcast news transcription, sections in article or books, and individual messages in a discussion board or an email trail. For non-version controlled documents finding the text segments is equivalent to finding the boundaries or edges between consecutive segments. See \cite{Hearst1997,Beeferman1999,Mccallum2000} for several recent studies in this area.

Things get a bit more complicated in the case of version controlled documents. Segments, and their boundaries exist in each version. As in case of image processing, we may view segment boundaries as edges in the space-time domain $\Omega$. These boundaries separate the segments from each other, much like borders separate countries in a two dimensional geographical map.

Assuming all edges are correctly identified, we can easily identify the segments as the interior points of the closed boundaries. In general, however, attempts to identify segment boundaries or edges will only be partially successful. As a result predicted edges in practice are not closed and do not lead to interior segments. We consider now the task of predicting segment boundaries or edges in $\Omega$ and postpone the task of predicting a segmentation to the next section.

Edges, or transitions between segments, correspond to abrupt changes in the local word distribution. We thus characterize them as points in $\Omega$ having high gradient value. In particular, we distinguish between vertical edges (transitions across document positions), horizontal edges (transitions across versions), and diagonal edges (transitions across both document position and version). These three types of edges may be diagnosed based on the magnitudes of $\dot\gamma_s$, $\dot\gamma_t$, and $\dot\alpha_1\gamma_s+\dot\alpha_2\gamma_t$ respectively.

\subsection{Experiments}\label{sec:edgeExp}

\begin{figure*} \centering
{ \begin{tabular}{ l  l l l  c c c  c c c  }
\toprule
\small \bf{Article} & \small  \bf{Rev.} & \small  \bf{Voc.}& \small $\pmb{p(y)}$ & \multicolumn{3}{ c }{ \small  \bf{Error Rate}} &  \multicolumn{3}{c}{ \small \bf{F1 Measure}} \\
& \small  				 & \small \bf{Size}	  & \small        & \small  a     & \small  b     & \small  c     & \small  a     & \small  b     & \small  c    \\
\midrule
\small Atlanta  			& \small  2000 & \small  3078 & \small  0.401 & \small  0.401 & \small  0.424 & \small  0.339 & \small  0.000 & \small  0.467 & \small  0.504\\
\small Religion 			& \small  2000 & \small  2880 & \small  0.403 & \small  0.404 & \small  0.432 & \small  0.357 & \small  0.000 & \small  0.470 & \small  0.552\\
\small Language 			& \small  2000 & \small  3727 & \small  0.292 & \small  0.292 & \small  0.450 & \small  0.298 & \small  0.000 & \small  0.379 & \small  0.091\\
\small European Union 	& \small  2000 & \small  2382 & \small  0.534 & \small  0.467 & \small  0.544 & \small  0.435 & \small  0.696 & \small  0.397 & \small  0.663\\
\small Beijing             & \small  2000 & \small  3857 & \small  0.543 & \small  0.456 & \small  0.474 & \small  0.391 & \small  0.704 & \small  0.512 & \small  0.682\\
\small Amazon Kindle FAQ   & \small  100  & \small   573 & \small  0.339 & \small  0.338 & \small  0.522 & \small  0.313 & \small  0.000 & \small  0.436 & \small  0.558\\
\bottomrule
\end{tabular}}
\caption{\small Test set error rate and F1 measure for edge prediction (section boundaries in Wikipedia articles and author change in Google Wave). The space-time domain $\Omega$ was divided to a grid with each cell labeled edge ($y=1$) or no edge ($y=0$) depending on whether it contained any edges. Method a corresponds to a predictor that always selects the majority class. Method b corresponds to the TextTiling test segmentation algorithm \cite{Hearst1997} without paragraph boundaries information. Method c corresponds to a logistic regression classifier whose feature set is composed of statistical summaries (mean, median, max, min) of $\dot\gamma_s(s,t)$ within the grid cell in question as well as neighboring cells. } \label{fig:edgePredictQuan}
\end{figure*}

Besides the synthetic data results in Figure~\ref{fig:religionPartVis}, we conducted edge detection experiments on six different real world datasets. Five datasets are Wikipedia.com articles: Atlanta, Religion, Language, European Union, and Beijing. Religion and European Union are version controlled documents with relatively frequent updates, while Atlanta, language, and Beijing have less frequent changes. The sixth dataset is the Google Wave Amazon Kindle FAQ which is a less structured version controlled document.

Preprocessing included removing html tags and pictures, word stemming, stop-word removal, and removing any non alphabetic characters (numbers and punctuations). The section heading information of Wikipedia and the information of author of each posting in Google Wave is used as ground truth for segment boundaries. This information was separated from the dataset and was used for training  and evaluation (on testing set).

Figure~\ref{fig:edgeVis} displays a gradient information, local maxima, and ground truth segment boundaries for the version controlled Wikipedia articles Religion and Atlanta. The local gradient maxima nicely match the segment boundaries which lead us to consider training a logistic regression classifier on a feature set composed of gradient value statistics (min, max, mean, median of $\|\dot\gamma_s(s,t)\|$ in the appropriate location as well as its neighbors (the space-time domain $\Omega$ was divided into a finite grid where each cell either contained an edge ($y=1$) or did not ($y=0$)). The table in Figure~\ref{fig:edgePredictQuan} displays the test set accuracy and F1 measure of three predictors: our logistic regression (method c) as well as two baselines: predicting edge/no-edge based on the marginal $p(y)$ distribution (method a) and  TextTiling (method b) \cite{Hearst1997} which is a popular text segmentation algorithm. Since we do not assume paragraph information in our experiment we ignored this component and considered the document as a sequence with $w=20$ and 29 minimum depth gaps parameters (see \cite{Hearst1997}). We conclude from the figure that the gradient information leads to better prediction than TextTiling (on both accuracy and F1 measure).

\section{Segmentation}

As mentioned in the previous section, predicting edges may not result in closed boundaries. It is possible to analyze the location and direction of the predicted edges and aggregate them into a sequence of closed boundaries surrounding the segments. We take a different approach and partition points in $\Omega$ to $k$ distinct values or segments based on local word content and space-time proximity.

For two points $(s_1,t_2),(s_2,t_2)\in\Omega$ to be in the same segment we expect $\gamma(s_1,t_1)$ to be similar to $\gamma(s_2,t_2)$ and for $(s_1,t_1)$ to be close to $(s_2,t_2)$. The first condition asserts that the two locations discuss the same topic. The second condition asserts that the two locations are not too far from each other in the space time domain. More specifically, we propose to segment $\Omega$ by clustering its points based on the following geometry
\begin{multline} \label{eq:segMetric}
d((s_1,t_1),(s_2,t_2)) = d_H(\gamma(s_1,t_1),\gamma(s_2,t_2))\\
+ \sqrt{c_1(s_1-s_2)^2 + c_2(t_1-t_2)^2}
\end{multline}
where $d_H:\mathbb{P}_V\times\mathbb{P}_V\to\mathbb{R}$ is Hellinger distance
\begin{align}
d_H^2(u,v) &= \sum_{i=1}^V (\sqrt{u_i}-\sqrt{v_i})^2.
\end{align}
The weights $c_1,c_2$ are used to balance the contributions of word content similarity with the similarity in time and space.

\begin{figure*} \centering
\setlength\fboxsep{0pt} \setlength\fboxrule{0.5pt}
\fbox{\includegraphics[width=0.31\textwidth]{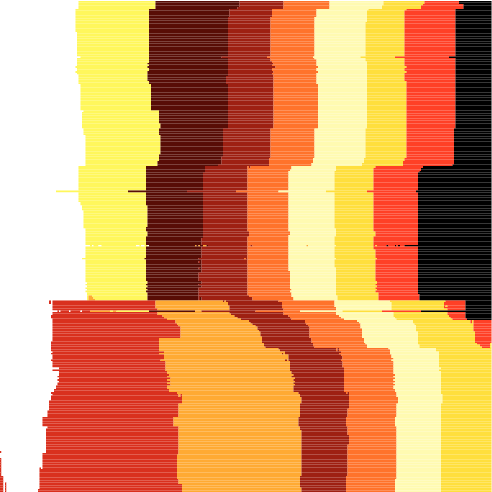}}
\fbox{\includegraphics[width=0.31\textwidth]{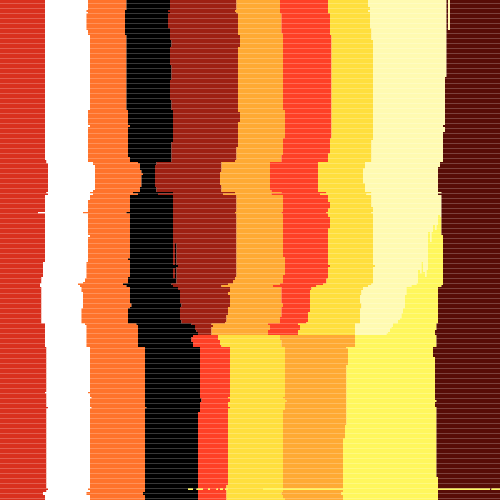}}
\fbox{\includegraphics[width=0.31\textwidth]{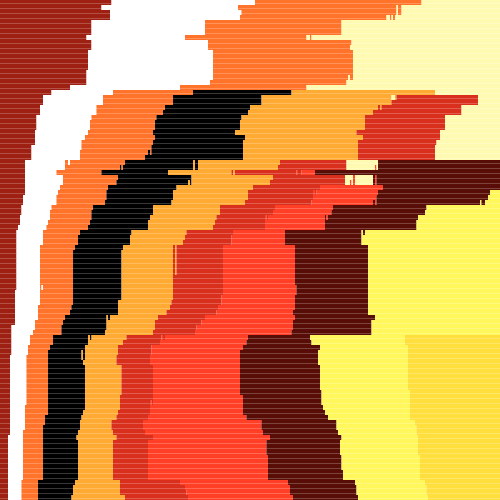}}
\fbox{\includegraphics[width=0.31\textwidth,height=0.31\textwidth]{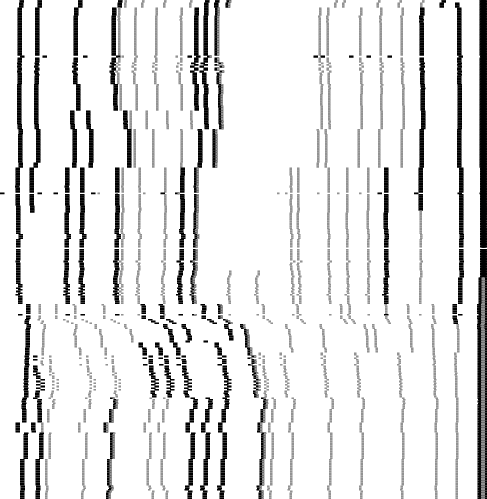}}
\fbox{\includegraphics[width=0.31\textwidth,height=0.31\textwidth]{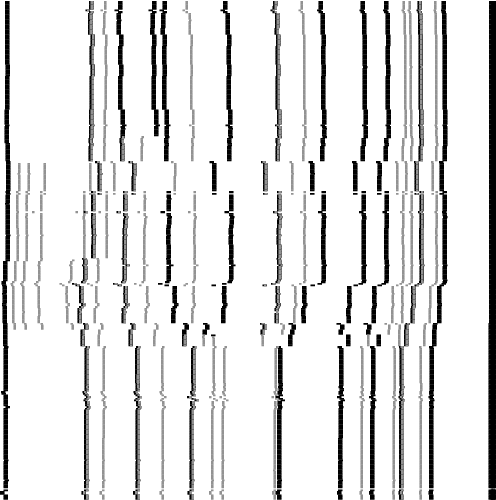}}
\fbox{\includegraphics[width=0.31\textwidth,height=0.31\textwidth]{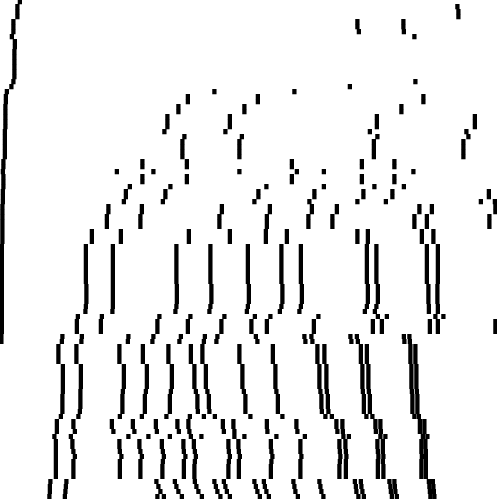}}\vspace{-5pt}
\caption{ \small Predicted segmentation (top) and ground truth segment boundaries (bottom) of portions of the version controlled Wikipedia articles  Religion (left), Atlanta (middle) and the Google Wave Amazon Kindle FAQ(right). The predicted segments match the ground truth segment boundaries. Note that the first 100 revisions are used in Google Wave result. The proportion of the segments that appeared in the beginning is keep decreasing while the revisions increases and new segments appears.} \label{fig:segVis}  \vspace{-9pt}
\end{figure*}

\subsection{Experiments}

Figure~\ref{fig:segVis} displays the ground truth segment boundaries and the segmentation results obtained by applying $k$-means clustering ($k=11$) to the metric \eqref{eq:segMetric}. The figure shows that the predicted segments largely match actual edges in the documents even though no edge or gradient information was used in the segmentation process.

\section{Predicting Future Operations}
The fourth and final task is predicting a future revision $d_{l+1}$ based on the smoothed representation of the present and past versions $d_1,\ldots,d_l$. In terms of $\Omega$, this means predicting features associated with $\gamma(s,t), t\geq t'$ based on $\gamma(s,t), t<t'$.

\subsection{Experiments}
We concentrate on predicting whether Wikipedia edits are reversed in the next revision. This action, marked by a label UNDO or REVERT in the Wikipedia API, is important for preventing content abuse or removing immature content (by predicting ahead of time suspicious revisions).

We predict whether a version will undergo UNDO in the next version using a support vector machine based on statistical summaries (mean, median, min, max) of the following feature set
$\|\dot\gamma_s(s,t)\|$, $\|\ddot\gamma_s(s,t)\|$, $\|\dot\gamma_t(s,t)\|)$, $\|\dot\gamma_t(s,t)\|$, $g(h)$, and $h(s)$. Figure~\ref{fig:futurePredictQuan} shows the test set error and F1 measure for the logistic regression based on the smoothed space-time representation (method c), as well as two baselines. The first baseline (method a) predicts the majority class and the second baseline (method b) is a logistic regression based on the term frequency content of the current test version. Using the derivatives of $\gamma$, we obtain a prediction that is better than choosing majority class or logistic regression based on word content. We thus conclude that the derivatives above provide more useful information (resulting in lower error and higher F1) for predicting future operations than word content features.

\begin{figure*}
\centering
\begin{tabular}{ l  l l l  c c c c c c c c }
\toprule
\small \bf{Article} & \small  \bf{Rev.} & \small  \bf{Voc.}& \small $\pmb{p(y)}$ &   \multicolumn{3}{ c }{\small \bf{Error Rate}} &  \multicolumn{3}{c}{\small \bf{F1 Measure}} \\
& \small  				 & \small \bf{Size} & \small 				    & \small    a   & \small  b     & \small  c     & \small  a     & \small  b     & \small  c    \\
\midrule
\small Atlanta  			& \small  2000 & \small  3078 & \small  	0.218   & \small  0.219 & \small  0.313 & \small  0.212 & \small  0.000 & \small  0.320 & \small  0.477\\
\small Religion 			& \small  2000 & \small  2880 & \small  	0.123   & \small  0.122 & \small  0.223 & \small  0.125 & \small  0.000 & \small  0.294 & \small  0.281\\
\small Language 			& \small  2000 & \small  3727 & \small 	0.189   & \small  0.189 & \small  0.259 & \small  0.187 & \small  0.000 & \small  0.334 & \small  0.455\\
\small European Union 	& \small  2000 & \small  2382 & \small 	0.213   & \small  0.208 & \small  0.331 & \small  0.209 & \small  0.000 & \small  0.275 & \small  0.410\\
\small Beijing             & \small  2000 & \small  3857 & \small      0.137   & \small  0.137 & \small  0.219 & \small  0.136 & \small  0.000 & \small  0.247 & \small  0.284\\
\bottomrule
\end{tabular}
\caption{ \small Error rate and F1 measure over held out test set of predicting future UNDO operation in Wikipedia articles. Method a corresponds to a predictor that always selects the majority class. Method b corresponds to a logistic regression based on the term frequency vector of the current version. Method c corresponds a logistic regression that uses summaries (mean, median, max, min) of $\|\dot\gamma_s(s,t)\|$, $\|\dot\gamma_s(s,t)\|$, $g(t)$, and $h(s)$.}
\label{fig:futurePredictQuan}\vspace{-5px}
\end{figure*}

\section{Related Work}
While document analysis is a very active research area, there has been relatively little work on examining version controlled documents. Our approach is the first to consider version controlled documents as continuous mappings from a space-time domain to the space of local word distributions. It extends the ideas in \cite{Lebanon2007b} of using kernel smoothing to create a continuous representation of documents. In fact, our framework generalizes  \cite{Lebanon2007b} as it reverts to it in the case of a single revision.

Other approaches to sequential analysis of documents concentrate on discrete spaces and discrete models, with the possible extension of \cite{Blei2009}. Related papers on segmentation and sequential document analysis are \cite{Hearst1997,Beeferman1999,Mccallum2000} with \cite{Hearst1997} being the closest in spirit to our approach.
An influential model for topic modeling within and across documents is latent Dirichlet allocation \cite{Blei2003,Blei2006}. Our approach differs in being fully non-parametric and in that it does not require iterative parametric estimation or integration. The interpretation of local word smoothing as a non-parametric statistical estimator \cite{Lebanon2007b} may be extended to our paper in a straightforward manner.

Several attempts have been made to visualize themes and topics in documents, either by keeping track of the word distribution or by dimensionality reduction techniques e.g.,   \cite{Fortuna2005,Havre2002,Spoerri1993,nvacbook}. Such studies tend to visualize a corpus of unrelated documents as opposed to ordered collections of revisions which we explore.

\section{Summary and Discussion}

The task of analyzing and visualizing version controlled document is an important one. It allows external control and monitoring of collaboratively authored resources such as Wikipedia, Google Wave, and CVS or SVN documents. Our framework is the first to develop analysis and visualization tools in this setting. It presents a new representation for version controlled documents that uses local smoothing to map a space-time domain $\Omega$ to the simplex of tf vectors $\mathbb{P}_V$. We demonstrate the applicability of the representation for four tasks: visualizing change, predicting edges, segmentation, and predicting future revision operations.

Visualizing changes may highlight significant structural changes for the benefit of users and help the collaborative authoring process. Improved edge prediction and text segmentation may assist in discovering structural or semantic changes and their evolution with the authoring process. Predicting future operation may assist authors as well as prevent abuse in coauthoring projects such as Wikipedia.

The experiments described in this paper were conducted on synthetic, Wikipedia and Google Wave articles. They show that the proposed formalism achieves good performance both qualitatively and quantitatively as compared to standard baseline algorithms.

It is intriguing to consider the similarity between our representation and image processing. Predicting segment boundaries are similar to edge detection in images. Segmenting version controlled documents may be reduced to image segmentation. Predicting future operations is similar to completing image parts based on the remaining pixels and a statistical model. Due to its long and successful history, image processing is a good candidate for providing useful tools for version controlled document analysis. Our framework facilitates this analogy and we believe is likely to result in novel models and analysis tools inspired by current image processing paradigms. A few potential examples are wavelet filtering, image compression, and statistical models such as Markov random fields.

\section*{Acknowledgements}
The research described in this paper was funded in part by NSF grant IIS-0746853.

\newpage
\bibliographystyle{coling}
\bibliography{../../share/externalPapers,../../share/groupPapers}
\end{document}